\pdfoutput=1
\documentclass[10pt,conference]{IEEEtran}
\IEEEoverridecommandlockouts

\newif\ifanonymous
\newif\ifshowacknowledgment

\anonymousfalse
\showacknowledgmenttrue

\def\BibTeX{{\rm B\kern-.05em{\sc i\kern-.025em b}\kern-.08em
    T\kern-.1667em\lower.7ex\hbox{E}\kern-.125emX}}

\makeatletter
\def\@IEEEsectpunct{\ }
\makeatother

\makeatletter
\renewcommand\paragraph{%
  \@startsection{paragraph}{4}{\parindent}%
  {0pt}%
  {0pt}%
  {\normalfont\normalsize\itshape}%
}
\makeatother

\usepackage{cite}
\usepackage{microtype}
\usepackage{amssymb,amsfonts}
\usepackage{algorithmic}
\usepackage{textcomp}
\usepackage{url}
\usepackage{etoolbox}
\usepackage[inline]{enumitem}

\usepackage[ruled,vlined,linesnumbered]{algorithm2e}
\SetArgSty{textrm}
\SetKwComment{tcp}{$\rhd\;$}{}%
\makeatletter
\patchcmd{\@algocf@start}%
{-1.5em}%
{0pt}%
{}{}%
\makeatother

\makeatother
\usepackage{threeparttable}
\usepackage{todonotes,xspace}
\usepackage[dvipsnames]{xcolor}
\definecolor{shadecolor}{rgb}{0.90, 0.90, 0.90}

\usepackage{stackengine} %

\setstackgap{L}{1.4\normalbaselineskip} %
\stackMath

\usepackage[many]{tcolorbox}

\definecolor{promptbg}{RGB}{245,245,245}    %
\definecolor{titlebg}{RGB}{235,235,235}      %

\newtcolorbox{promptbox}[1]{
    enhanced,
    breakable,
    colback=promptbg,        %
    colbacktitle=titlebg,    %
    colframe=titlebg,
    coltitle=black,          %
    fonttitle=\bfseries\text\small,
    fontupper=\text\small,
    title=#1,               %
    boxrule=1.5pt,          %
    arc=1pt,                %
    left=10pt,              %
    right=10pt,             %
    top=4pt,                %
    bottom=2pt,             %
    toptitle=1pt,           %
    bottomtitle=1pt,        %
    titlerule=0.5pt,        %
    titlerule style={titlebg} %
}

\usepackage{graphicx}
\usepackage{caption}
\usepackage{subcaption}
\usepackage{amsmath}
\usepackage{thm-restate}

\usepackage{circledsteps}

\renewcommand{\Circled}[1]{\textcircled{\raisebox{-.87pt}{\texttt{#1}}}}

\usepackage{mathtools}
\usepackage{csquotes}

\usepackage{fancyvrb}
\usepackage[frozencache,cachedir=minted-cache]{minted} %

\setminted[c]{
    frame=lines,
    framesep=2mm,
    baselinestretch=1.0,
    fontsize=\footnotesize, %
    breaklines=true,
    xleftmargin=0em,
    xrightmargin=0em,
    autogobble,
    escapeinside=||, %
    showspaces=false,
    showtabs=false
}
\newcommand{\old}{\ifmmode\mathtt{old}\else\texttt{old}\fi}
\newcommand{\valid}{\ifmmode\mathtt{valid}\else\texttt{valid}\fi}

\usepackage{framed}

\usepackage{adjustbox}
\usepackage{makecell}
\usepackage{booktabs} %
\usepackage{tikz}
\usetikzlibrary{patterns,calc,arrows,shapes,snakes,automata,backgrounds,petri,positioning,decorations.pathmorphing,intersections}
\usepackage{tkz-euclide}
\tikzset{mynode/.style={draw,solid,circle,inner sep=1pt}}
\usepackage[customcolors]{hf-tikz}
\hfsetfillcolor{gray!10}
\hfsetbordercolor{none}
\usepackage{pgfplots}
\usepgfplotslibrary{units}
\usepgfplotslibrary{groupplots}
\usepgfplotslibrary{fillbetween}
\pgfplotsset{compat=1.17}
\usepackage{multirow}
\usepackage{pifont} 

\usepackage{hyperref}
\hypersetup{
	colorlinks=true,
	linkcolor=RedViolet,
	anchorcolor=black,
	citecolor=blue,
	urlcolor=black,
	breaklinks=true
}

\usepackage[
    type={CC},
    modifier={by-nc-sa},
    version={3.0},
]{doclicense}

\usepackage[capitalise,noabbrev]{cleveref}

\crefname{equation}{}{}
\crefname{figure}{Fig.}{Figs.}

\crefdefaultlabelformat{#2\textup{#1}#3}

\newcommand{\rte}{\textup{\textsc{Frama-C/Rte}}\xspace}
\newcommand{\Wp}{\textup{\textsc{Frama-C/Wp}}\xspace}
\newcommand{\framac}{\textup{\textsc{Frama-C}}\xspace}

\newcommand{\preguss}{\textsc{Preguss}\xspace}
\newcommand{\verdict}{\textsc{Verdict}\xspace}

\makeatletter
\newcommand{\downbrace}[1]{%
  \mathop{\vtop{\m@th\ialign{##\crcr
    $\hfil\displaystyle{#1}\hfil$\crcr
    \noalign{\kern3\p@\nointerlineskip}%
    \downbracefill\crcr
    \noalign{\kern3\p@}\crcr}}}\limits}
\makeatother
\newcommand{\eeq}{~{}={}~}

\def\triangleforqed{\hbox{$\lhd$}}
\makeatletter
\DeclareRobustCommand{\qedT}{%
	\ifmmode
	\eqno \def\@badmath{$$}%
	\let\eqno\relax \let\leqno\relax \let\veqno\relax
	\hbox{\triangleforqed}%
	\else
	\leavevmode\unskip\penalty9999 \hbox{}\nobreak\hfill
	\quad\hbox{\triangleforqed}%
	\fi
}
\makeatother

\usepackage{amsthm}
\theoremstyle{plain}
\newtheorem*{thm*}{Theorem}
\newtheorem*{exmp*}{Example}
\newtheorem*{counterexmp*}{Counterexample}

\newtheoremstyle{ourstyle}{}{}{\itshape}{}{\bfseries}{.}{ }{\thmname{#1}\thmnumber{ #2}\thmnote{ (#3)}}
\theoremstyle{ourstyle}

\newtheoremstyle{exmpstyle}{}{}{}{}{\bfseries}{.}{ }{\thmname{#1}\thmnumber{ #2}\thmnote{ (#3)}}
\theoremstyle{exmpstyle}

\newtheoremstyle{rmkstyle}{}{}{}{}{\bfseries}{.}{ }{\thmname{#1}\thmnote{ (#3)}}
\theoremstyle{rmkstyle}

\newcounter{challenge}
\renewcommand{\thechallenge}{\arabic{challenge}} %

\crefname{challenge}{Challenge}{Challenges}   %
\Crefname{challenge}{Challenge}{Challenges}   %

{%
  \refstepcounter{challenge}%
  \begin{tcolorbox}[boxrule=1pt,colback=white,colframe=black!75,boxsep=0mm,#1]%
  \textbf{Challenge \thechallenge.}\ %
}
{%
  \end{tcolorbox}%
}

\newcounter{insight}
\renewcommand{\theinsight}{\arabic{insight}} %

\crefname{insight}{Insight}{Insights}   %
\Crefname{insight}{Insight}{Insights}   %

{%
  \refstepcounter{insight}%
  \begin{tcolorbox}[boxrule=1pt,colback=white,colframe=black!75,boxsep=0mm,#1]%
  \textbf{Insight \theinsight.}\ %
}
{%
  \end{tcolorbox}%
}

\newcommand{\NEWCITE}{}

\usepackage{tabularx}

\newcommand{\agentdetails}{See \cref{app:agent} for details.}

\begin{document}

\title{\verdict: Multi-Agent Parallel Modular Verification
}

\author{%
\IEEEauthorblockN{%
Tengjie Lin\textsuperscript{1},
Zhongyi Wang\textsuperscript{1},
Haokun Li\textsuperscript{2},
Jianwei Yin\textsuperscript{1}, and
Mingshuai Chen\textsuperscript{1,*}%
\thanks{\textsuperscript{*}The corresponding author.}}
\IEEEauthorblockA{%
\textsuperscript{1}Zhejiang University, Hangzhou, China \quad
\textsuperscript{2}Peking University, Beijing, China\\
\textsuperscript{1}\{tengjie.lin,zhongyi.wang,zjuyjw,m.chen\}@zju.edu.cn \quad
\textsuperscript{2}ker@pm.me}}

\maketitle
\bstctlcite{IEEEexample:BSTcontrol}%

\begin{abstract}
Deductive verification scales to large software systems through \emph{modular reasoning}, where formal contracts summarize the behavior of individual functions and enable their independent verification. Its central bottleneck is \emph{specification synthesis}: constructing the contracts required for modular verification. Recent advances in large language models (LLMs) have enabled automated specification synthesis at unprecedented scale. However, existing approaches remain fundamentally \emph{bottom-up}, requiring callees to be verified before their callers. This dependency limits parallelism and provides no principled mechanism for resolving interprocedural inconsistencies that propagate across call chains.

We present \verdict, a multi-agent framework for parallelizing modular deductive verification of large-scale programs. \verdict is built on two key principles. First, \emph{assumption-based parallel specification synthesis} assigns an agent to each function and allows it to proceed immediately using assumed contracts for its callees, enabling verification to proceed in parallel across the program. Second, \emph{bidirectional interprocedural conflict reconciliation} resolves the inconsistencies introduced by these assumptions: when a synthesized contract contradicts an assumption, agents propagate the necessary revisions both up and down the call graph until the contracts converge. Experimental results on real-world C programs show that \verdict synthesizes higher-quality specifications than existing approaches while substantially reducing verification time. In particular, \verdict is the first approach that scales to the complete 20,000-line X.509 parser, ruling out nearly all the potential runtime errors within hours, whereas prior tools like \preguss fall far short even after days of execution.
\end{abstract}


\section{Introduction}\label{sec:introduction}

Automated verification of large-scale software is 
an essential yet 
long-standing challenge in both formal methods and software engineering. A key enabler for modern deductive verifiers to scale to real-world systems is \emph{modularization}~\cite{deductive-verification}: Each module (typically a function) is associated with a \emph{contract}---%
comprising preconditions and postconditions---that captures the semantic behavior of the function. Once a function is verified against its contract, every call to it simply invokes the contract rather than re-examining its implementation. Consequently, the primary human effort in verifying a large-scale program boils down to \emph{specification synthesis}~\cite{specification-synthesis}, namely, to construct both function contracts and auxiliary intraprocedural annotations, such as loop invariants and assertions. Crafting these comprehensive formal specifications for large-scale programs with non-trivial call hierarchies is notoriously difficult and has long relied on deep expertise~\cite{fm21-jcvm-verification,x509-parser}.

Recent efforts---notably {\preguss}~\cite{preguss} and SpecSyn~\cite{SpecSyn}---approach the problem by leveraging large language models (LLMs) to \emph{automate} the synthesis of specifications through a \emph{divide-and-conquer} strategy: On the ``divide'' side, the entire program is decomposed into a sequence of verification units (V-Units), each containing only a small set of functions.
By restricting each LLM query to such a localized context, they effectively sidestep the \emph{long-context reasoning bottleneck}~\cite{long-context-reasoning-1,long-context-reasoning-2}---LLMs' well-known tendency to lose relevant information and degrade as their input context grows, due to attention dispersion~\cite{attention-despersion-1,attention-despersion-2}. On the ``conquer'' side, these V-Units are validated in a \emph{bottom-up} order~\cite{cav24-autospec,preguss,SpecSyn}, i.e., callees before callers, which is a natural extension of the modularization principle of deductive verification (see~\cref{sec-sub:bg:bottom-up-verification-paradigm}).

Despite its widespread adoption, this bottom-up paradigm exhibits two fundamental limitations. 
\begin{enumerate*}[label=(\Roman*)]
\item \emph{Restricted parallelism}: While units at the same call level can be verified in parallel, units across different levels must follow a strict callees-before-callers order. Consequently, the verification of higher-level callers remains idle until all lower-level callees have been verified, thus leaving substantial parallelism unexploited.\label{challenge:parallelism}
\item The \emph{absence of a principled mechanism for reconciling interprocedural conflicts}: The verification of a caller may require a weaker precondition or a stronger postcondition of a callee, while modifications to a callee's contract may invalidate the caller's proof obligations, thus forcing the revision of the caller's own contract accordingly. Resolving such conflicts often requires \emph{bidirectional} propagation of contract adjustments across the call graph. For instance, a change to a leaf function may ripple upward through all transitive callers, each of which must be re-verified and may, in turn, trigger further rounds of contract refinement. However, the statically partitioned and locally scoped verification units employed by traditional bottom-up approaches---each with visibility into only a bounded portion of the program---provide no systematic support for such dynamic, cross-unit coordination.\label{challenge:reconciling}
\end{enumerate*}
See \Cref{sec:motivation} for concrete examples illustrating the two limitations.

In response to the aforementioned challenges, we present {\verdict}, a multi-agent framework that essentially parallelizes the modular deductive verification of large-scale programs. At the core of {\verdict} are two key architectural components:
\begin{enumerate*}[label=(\roman*)]
\item \emph{a parallel specification-synthesis mechanism} and 
\item \emph{an interprocedural conflict-reconciliation protocol}.
\end{enumerate*}
The former addresses Challenge \ref{challenge:parallelism}: {\verdict} abandons the sequential bottom-up workflow and instead assigns each function a persistent \emph{agent} responsible for maintaining its contract. Rather than waiting for callees to publish their specifications, agents proceed by \emph{assuming contracts} for unresolved callees (\cref{sec-sub:assumption-based-verification}), thereby enabling concurrent specification synthesis across all levels of the call graph. 
Our second component addresses Challenge \ref{challenge:reconciling}: {\verdict} coordinates agents through a typed event protocol. A \emph{Supply} event propagates a callee's contract change upward to its callers, while a \emph{Demand} event propagates a caller's specification request downward to its callees (\cref{sec-sub:event-driven-incremental}). These events cascade bidirectionally along the call chains, and thus provide a conflict-reconciliation mechanism across arbitrary call distances.

We have implemented {\verdict} on top of the Claude Agent SDK \cite{claude-agent-sdk} with {\Wp}  \cite{frama-c/wp} as the underlying deductive verifier.  %
{\Wp} certifies that a C program satisfies a certain
program property, notably, the absence of \emph{runtime errors} (RTEs) such as invalid memory access or division by zero~\cite{c-standard}. We focus on programs without recursive data structures that require inductive predicates (\cref{sec:limitations}).
Evaluation on 15 real-world benchmarks reveals that {\verdict} substantially outperforms state-of-the-art baselines, {\preguss} and SpecSyn, in both specification quality and verification time. In particular, increasing the number of agents improves efficiency, while specification quality converges within a small refinement budget, indicating that the two objectives improve together rather than trade off.
Moreover, {\verdict} enables almost fully automated RTE-freeness verification of the complete X.509 certificate parser, with over 20{,}000 lines of code (LoC), in a few hours---a task that previously required months of expert manual effort~\cite{x509-parser}.

\smallskip
\noindent\textbf{Contributions.}
Our main contributions are as follows.
\begin{itemize}
    \item We identify two fundamental limitations of the divide-and-conquer, bottom-up paradigm for LLM-assisted specification synthesis for large-scale real-world programs. 
    \item We present {\verdict}, a multi-agent framework that overcomes these limitations by replacing sequential bottom-up synthesis with assumption-based parallel dispatch and event-driven bidirectional reconciliation. %
    \item We design verification agents that run an adaptive loop with domain-specific tools for verification and cross-agent coordination.
    \item We implement {\verdict} and
    show that it substantially outperforms state-of-the-art baselines in both specification quality and efficiency. {\verdict} is, to the best of our knowledge, %
    the first automated approach capable of proving RTE-freeness for real-world programs with over 20,000 LoC, which has previously been limited to programs of roughly 1,000 LoC (cf.~\preguss~\cite{preguss}).
\end{itemize}
\section{Background}\label{sec:background}

\begin{figure}[t]
	\centering
    \includegraphics[width=\linewidth]{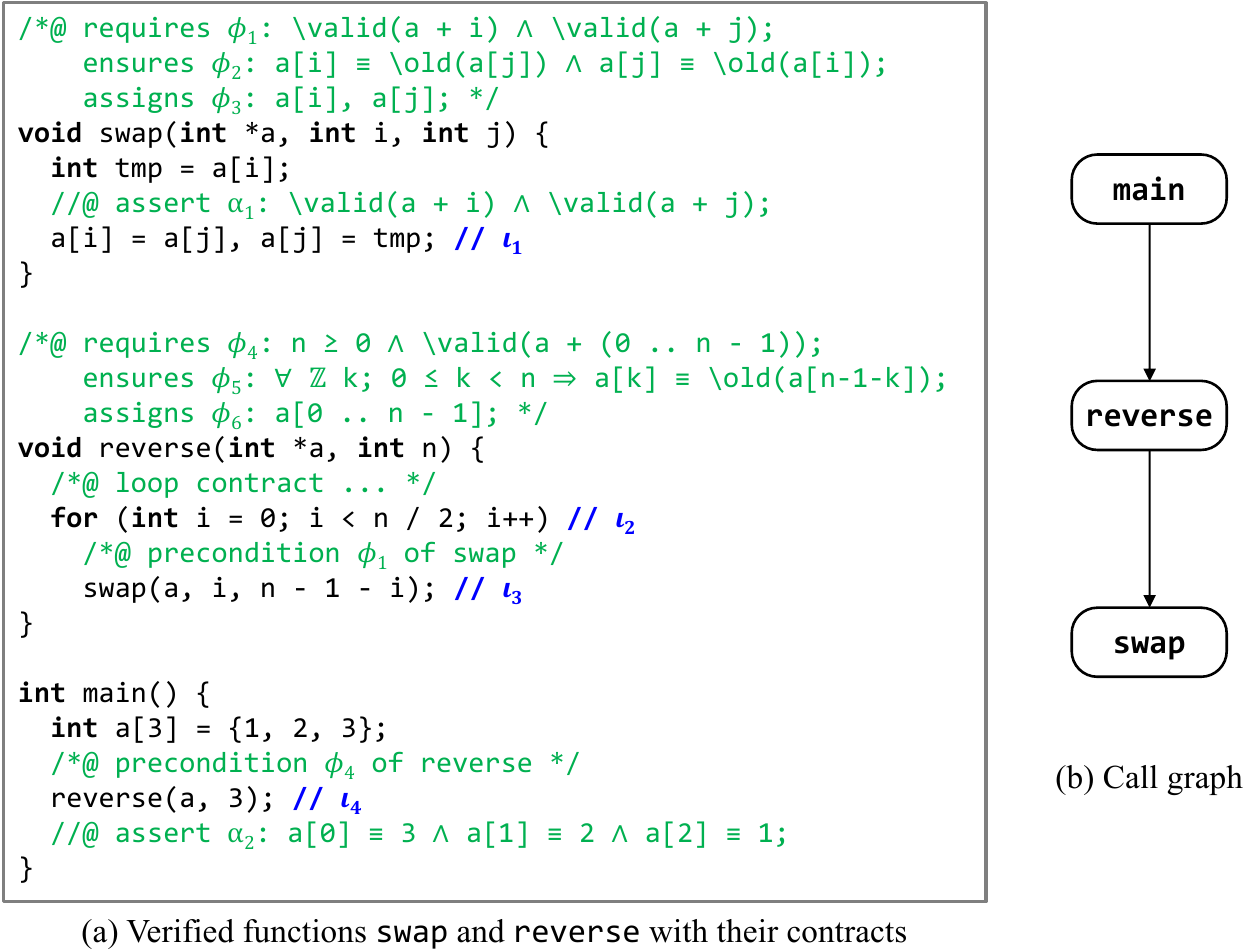}
    \caption{Illustration of modular program verification.}
	\label{fig:example-1}
\end{figure}

\subsection{Modular Verification}\label{sec-sub:bg:modular-verification}

Modularization is the common wisdom for handling large-scale software systems with increasing complexity~\cite[Chapt. 9]{deductive-verification}. In deductive verification, modularization is achieved through the \emph{design-by-contract} principle~\cite{design-by-contract}: Each function is annotated with a contract that abstracts away its concrete implementation. %
Such a contract typically consists of \emph{preconditions}---the obligations that must be satisfied at the call site---and \emph{postconditions}---the guarantees established upon return. Verifying a function $f$ against its contract then amounts to proving the \emph{Hoare triple}
\[
\{P\}\;f\;\{Q\},
\]
where $P$ is the precondition, and $Q$ is the postcondition. The Hoare triple is valid iff every execution of $f$ starting from a state satisfying $P$ terminates in a state satisfying $Q$.

\Cref{fig:example-1} illustrates modular verification on a small C program. The function \texttt{swap} is annotated with a contract expressed in the ANSI/ISO C Specification Language (ACSL)~\cite{acsl} that (i)~requires the memory locations \texttt{a + i} and \texttt{a + j} to be validly writable (precondition $\phi_1$), and (ii)~ensures the values at these positions are exchanged (postcondition $\phi_2$), while also declaring that only these two locations are modified (frame condition $\phi_3$, a special case of postconditions). Once \texttt{swap} is verified against this contract, i.e., the deductive verifier (e.g., \Wp~\cite{frama-c/wp}) proves $\{\phi_1\}\;\texttt{swap}\;\{\phi_2\land\phi_3\}$, every \emph{call} to \texttt{swap} can soundly be replaced by its contract. 

The caller \texttt{reverse} is annotated with its own contract, which requires a valid array and a non-negative length ($\phi_4$), ensures the array is reversed ($\phi_5$), and declares the entire array as modified ($\phi_6$). Consequently, when verifying \texttt{reverse} (i.e., $\{\phi_4\}\;\texttt{reverse}\;\{\phi_5\land\phi_6\}$) which invokes \texttt{swap} at $\iota_3$ within a loop, the verifier need only check that the arguments at $\iota_3$ satisfy $\phi_1$, and can then assume $\phi_2$ and $\phi_3$ after the call, without ever inspecting \texttt{swap}'s body again. For brevity, we omit the loop contract (e.g., loop invariants, variants, and frame conditions) at $\iota_2$, though such intraprocedural annotations are necessary for the verifier to establish the contract of \texttt{reverse}.

\subsection{Proof Goals}\label{sec-sub:bg:proof-goals}

The purpose of establishing contracts is to prove program \emph{properties}, in our case, the absence of runtime errors~\cite{c-standard}.
A substantial portion of them can be encoded as first-order logic \emph{assertions}, such as $\alpha_1$ at $\iota_1$ and $\alpha_2$ after $\iota_4$ in \cref{fig:example-1}~(a). Validating $\alpha_1$---that \texttt{a + i} and \texttt{a + j} are valid for writing---depends on the precondition $\phi_1$ of \texttt{swap}; validating $\alpha_2$---that the array \texttt{a} has been reversed---depends on the postcondition $\phi_5$ and $\phi_6$ of \texttt{reverse}. We call these assertion-encoded properties---whether machine-generated~\cite{frama-c/rte} or human-crafted---\emph{proof goals}; a proof goal is discharged only when the assertion and all the specifications it depends on are jointly proved.

\subsection{Bottom-up Verification Paradigm}\label{sec-sub:bg:bottom-up-verification-paradigm}

A natural consequence of the design-by-contract principle (\cref{sec-sub:bg:modular-verification,sec-sub:bg:proof-goals}) is that \emph{callers cannot be fully verified until their callees' contracts are established}. This gives rise to the \emph{bottom-up} verification paradigm, widely adopted in recent LLM-based specification synthesis~\cite{cav24-autospec,preguss,SpecSyn}: specifications are constructed for leaf functions first, and verified contracts propagate upward along the call graph. In \cref{fig:example-1}, one must first verify \texttt{swap} (proving that it exchanges two array elements, $\phi_2$), then \texttt{reverse} (proving array reversal, $\phi_5$, using \texttt{swap}'s contract at $\iota_3$), and finally \texttt{main} (proving $\alpha_2$, which depends on $\phi_5$ and $\phi_6$). Moreover, satisfying callee preconditions is the primary driver for constructing caller preconditions: \texttt{reverse} imposes $\phi_4$ precisely to discharge $\phi_1$ of \texttt{swap} at $\iota_3$.
\section{Motivation}\label{sec:motivation}

\begin{figure}[t]
	\centering
    \includegraphics[width=\linewidth]{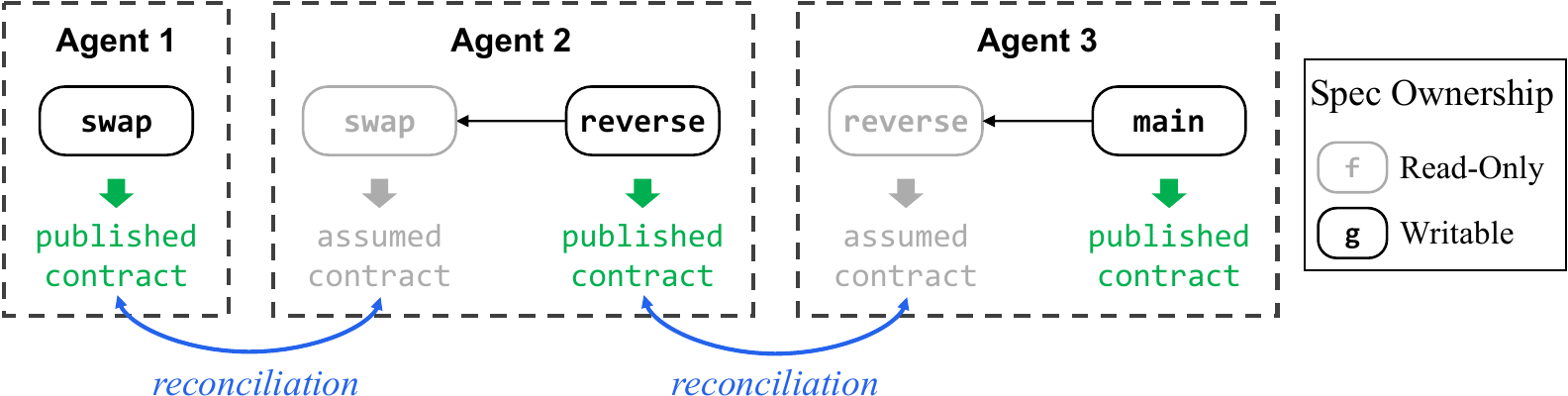}
    \caption{Assumption-based parallel specification synthesis for the program in \cref{fig:example-1}.}
	\label{fig:motivation-1}
\end{figure}

\begin{figure*}[t]
    \centering
    \includegraphics[width=\linewidth]{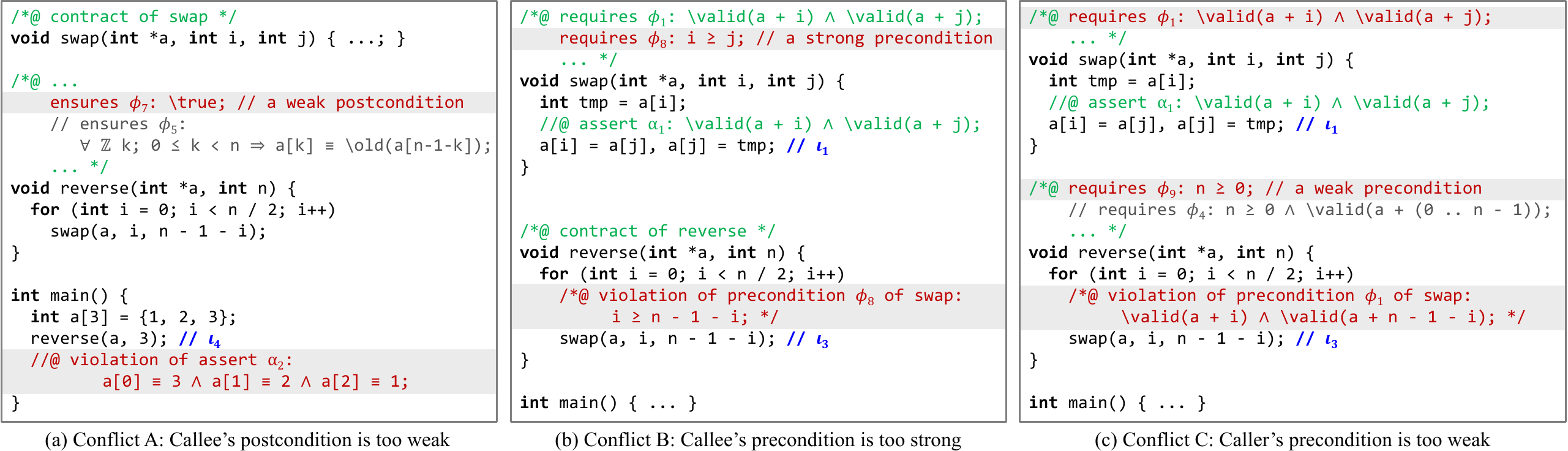}
    \caption{Three scenarios of interprocedural conflicts (highlighted by gray boxes) and their
    reconciliation, based on \cref{fig:example-1}.
    (a)~\texttt{main} demands that \texttt{reverse} strengthen its
    postcondition (replace $\phi_7$ with $\phi_5$).
    (b)~\texttt{reverse} demands that \texttt{swap} weaken its
    precondition (delete $\phi_8$).
    (c)~\texttt{reverse} strengthens its own precondition (replace
    $\phi_9$ with $\phi_4$), cascading upward to \texttt{main}.
    }
    \label{fig:motivation-2}
\end{figure*}

\subsection{Assumption-Based Parallel Specification Synthesis}\label{sec-sub:motiv:assumption-based-parallel-specsyn}

The bottom-up paradigm fundamentally limits parallelism. Consider \cref{fig:example-1}, verifying \texttt{swap} and \texttt{reverse} involves two time-consuming processes: LLM inference (generating and refining candidate specifications) and verifier validation ({\Wp} checking candidate specifications); these two processes are tightly interleaved within each verification unit's guess-and-check cycle~\cite{ASE24-Wu}. Moreover, the bottom-up ordering forces these cycles into a rigid sequence: \texttt{swap} must complete before \texttt{reverse} can begin, and \texttt{reverse} before \texttt{main}. Consequently, at any moment, only one level of the call graph actively progresses, while units at higher levels remain idle.

\verdict is motivated by a simple but consequential insight: \emph{since verification is modular---a function's proof depends only on its callees' contracts, not their implementations---we can begin verifying a caller under assumed contracts for its callees, without waiting for those callees' verification to finish.} In our approach, each function is assigned a dedicated Agent responsible for (i)~generating \emph{assumed contracts} for its callees from their call-site contexts and function bodies, (ii)~constructing its own contract and intraprocedural specifications and \emph{publishing} them, and (iii)~when a callee's agent later publishes a contract that conflicts with the assumed one, determining the appropriate \emph{reconciling} strategy (detailed in \cref{sec-sub:motiv:bid-interpro-conflict-reconcile}). \Cref{fig:motivation-1} illustrates this principle: the agents for \texttt{swap}, \texttt{reverse}, and \texttt{main} launch concurrently; each makes assumptions about its callees (e.g., the \texttt{reverse} agent assumes a contract for \texttt{swap}) and proceeds to verify. Because these agents operate independently, the system achieves substantial parallelism: given sufficient resources, every function in a large program can begin verification simultaneously.

\subsection{Bidirectional Interprocedural Conflicts Reconcilation}\label{sec-sub:motiv:bid-interpro-conflict-reconcile}

\begin{figure}[t]
	\centering
    \includegraphics[width=\linewidth]{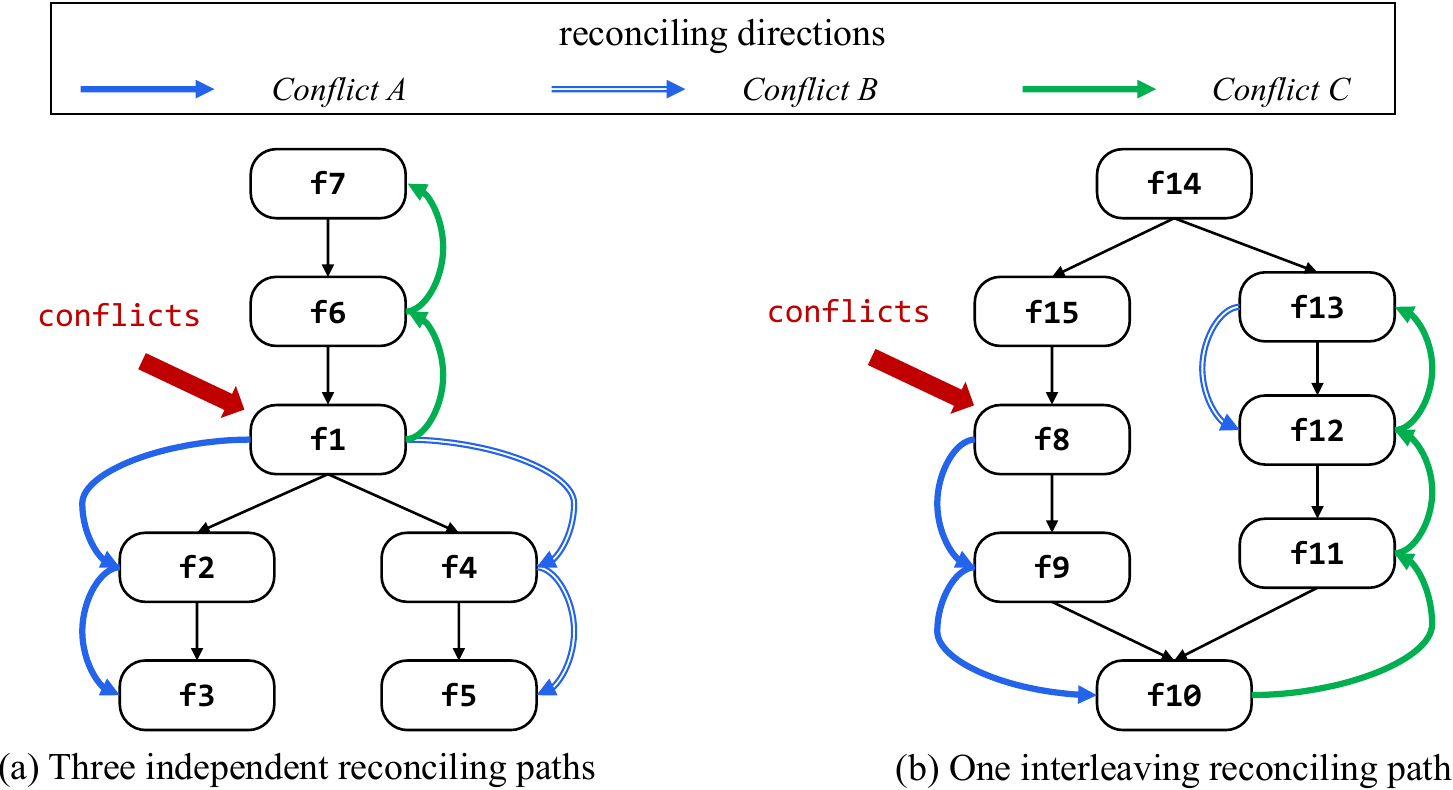}
    \caption{Reconciling directions of interprocedural conflicts. (a)~Three independent reconciling paths originating from conflicts at \texttt{f1}: downward postcondition strengthening demands (\texttt{f1}$\xrightarrow{A}$\texttt{f2}$\xrightarrow{A}$\texttt{f3}),
    downward precondition weakening demands (\texttt{f1}$\xrightarrow{B}$\texttt{f4}$\xrightarrow{B}$\texttt{f5}), and upward precondition strengthening cascades (\texttt{f1}$\xrightarrow{C}$\texttt{f6}$\xrightarrow{C}$\texttt{f7}). (b)~Interleaving of all three forms: a single conflict at \texttt{f8} triggers alternating downward and upward adjustments across the call chain.}
	\label{fig:motivation-2-propagation}
\end{figure}

As observed in recent work~\cite{preguss,icse25-specgen,SpecSyn}, LLMs rarely produce correct specifications in a single attempt for programs with non-trivial call hierarchies. These approaches therefore embed repair mechanisms that iteratively refine invalid specifications via verification feedback. Low-level errors---syntactic mistakes (e.g., misspelled variables) or missing intraprocedural annotations (e.g., loop invariants)---are localized and relatively easy to fix. The deeper challenge lies in reconciling \emph{interprocedural conflicts}: when a proof goal in one function cannot be discharged, the root cause often resides not in its own specification but in a deficient contract elsewhere in the call chain. We identify three canonical forms of interprocedural conflicts and construct representative scenarios for each by introducing intentional specification deficiencies into the example of \cref{fig:example-1}. \Cref{fig:motivation-2} summarizes these forms and their reconciling measures:

\emph{Conflict A: Callee's postcondition is too weak.}
Suppose \texttt{reverse} is initially published with only a trivial postcondition $\phi_7 \equiv \texttt{\textbackslash{}true}$ (\cref{fig:motivation-2}~(a)). When the agent for \texttt{main} attempts to discharge $\alpha_2$, it finds that $\phi_7$ provides no usable information. The only remedy is to \emph{demand} that \texttt{reverse} strengthen its postcondition to the full reversal formula $\phi_5$. This demand flows \emph{downward} along the call graph. In a deeper graph, such demands can propagate, as illustrated in \cref{fig:motivation-2-propagation}~(a): to prove $\phi_5$, \texttt{reverse} may need stronger postconditions from \texttt{swap} (e.g., that \texttt{swap} modifies exactly the two indexed elements), triggering a chain of downward refinements.%

\emph{Conflict B: Callee's precondition is too strong.}
Consider a variant in which \texttt{swap} carries a defensive precondition $\phi_8 \equiv \texttt{i} \ge \texttt{j}$ (\cref{fig:motivation-2}~(b)).  When \texttt{reverse} verifies its call site $\texttt{swap(a, i, n-1-i)}$ at $\iota_3$, it must prove $\phi_8[\texttt{i}\mapsto \texttt{i},\texttt{j}\mapsto \texttt{n-1-i}]$, i.e., $\texttt{i} \ge \texttt{n-1-i}$, which fails for every iteration where $\texttt{i} < \texttt{(n-1)/2}$.  The conflict is resolved by a \emph{precondition weakening demand}: \texttt{reverse} requests that \texttt{swap} remove $\phi_8$. Same as reconciling Conflict A, this demand flows from caller to callee (\cref{fig:motivation-2-propagation}~(a)); accepting it may force the callee to re-prove its internal goals under the weaker assumption, potentially cascading further down.

\emph{Conflict C: Caller's precondition is too weak.}
The third form arises when no change to a callee suffices---the caller must instead strengthen \emph{its own} contract (\cref{fig:motivation-2}~(c)).  Suppose \texttt{reverse} is initially published with only the weak precondition $\phi_9 \equiv \texttt{n} \ge \texttt{0}$, omitting memory validity. To prove the call-site precondition $\phi_1$ of \texttt{swap} at $\iota_3$, the verifier needs $\valid(\texttt{a}+\texttt{(0..n-1)})$, which $\phi_9$ does not supply. \texttt{reverse} must therefore strengthen its own precondition to $\phi_4 \equiv \texttt{n} \ge \texttt{0} \land \valid(\texttt{a}+\texttt{(0..n-1)})$.  This change cascades \emph{upward} (\cref{fig:motivation-2-propagation}~(a)): every caller of \texttt{reverse} must now establish $\phi_4$ at the call site, and if its own precondition is too weak to do so, the same conflict reappears one level higher---potentially cascading along an arbitrary prefix of the call chain.

\emph{Why do bottom-up approaches struggle at reconciling interprocedural conflicts?}
Crucially, the three conflict forms do not arise in isolation. \Cref{fig:motivation-2-propagation}~(b) illustrates the interleaving with an abstracted call chain: a Conflict~A demand propagates \texttt{f8}$\xrightarrow{A}$\texttt{f9}$\xrightarrow{A}$\texttt{f10},\footnote{We denote reconciling directions by labeled arrows: $\xrightarrow{X}$, where $X \in \{A, B, C\}$ denotes the type of the conflict.} strengthening \texttt{f10}'s postcondition; but to prove that stronger postcondition, \texttt{f10} must first strengthen its own precondition (Conflict~C), which cascades upward as \texttt{f10}$\xrightarrow{C}$\texttt{f11}$\xrightarrow{C}$\texttt{f12}$\xrightarrow{C}$\texttt{f13}; and if the new precondition is overly aggressive, it can trigger Conflict~B at a neighboring call site (\texttt{f13}$\xrightarrow{B}$\texttt{f12}). In short, the three forms \emph{interleave} along the call graph. Reconciling them thus requires two capabilities: (i)~the ability to modify contracts \emph{bidirectionally}---downward to callees (for Conflicts A and B) and upward to callers (for Conflict C)---in an interleaved manner; and (ii)~the ability to reach functions \emph{arbitrarily far} from the original conflict point, since a single initial deficiency can trigger adjustments across multiple call levels. The bottom-up, divide-and-conquer paradigm lacks both. First, it enforces a strictly upward flow: specifications are constructed for callees and then consumed by callers, with no mechanism for a caller to request refinement of an already-sealed callee contract---precisely what Conflict~A and~B require. Second, each verification unit exposes only a small, locally bounded set of functions; propagating a repair across arbitrary call levels is simply beyond what any such unit can express. These two limitations stem from the bottom-up design itself, not from any particular instantiation.

\begin{figure}[t]
    \centering
    \includegraphics[width=\linewidth]{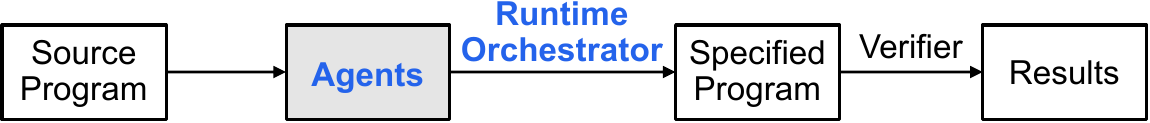}
    \caption{The workflow of {\verdict}.}
    \label{fig:overflow}
\end{figure}

\section{Overview of \verdict}\label{sec:overall}

{\verdict} overcomes the aforementioned limitations of the bottom-up paradigm with two key components, as depicted in \Cref{fig:overflow}. First, each function is assigned a persistent \emph{Agent} that owns its specification and can begin verification immediately under assumed contracts for unpublished callees.
Second, a \emph{Runtime Orchestrator} coordinates these agents through typed events---\emph{Event Supply} for upward cascades and \emph{Event Demand} for downward requests---and enforces a four-state machine per agent whose transitions are triggered exclusively by these events.
This design achieves parallelism without ordering constraints while providing the infrastructure for bidirectional conflict reconciliation.
We elaborate on the above two components in the subsequent sections, respectively.

\section{The Verification Agents}\label{sec:agent}

\subsection{Agent Dispatch}\label{sec-sub:agent-dispatch}

{\verdict} decomposes the input program into \emph{Verification Units} (V-Units), one per function. Each V-Unit bundles a \emph{host function} with its direct callees, where the host's contract and intraprocedural specifications are \emph{writable} by its assigned agent, while the callees' contracts are \emph{read-only} by the same agent. This \emph{exclusive} ownership model is essential: Since each function is owned by exactly one agent, any request to revise its contract is routed to that agent alone, which determines the guarantees the contract should provide, thus making cross-agent coordination traceable. Without exclusive ownership, multiple callers could modify the same contract concurrently, and thereby produce inconsistent revisions.

\subsection{Assumption-Based Verification}\label{sec-sub:assumption-based-verification}

An agent encounters a dilemma upon its first launch: its callees may not yet have published contracts, but waiting for them would forfeit the parallelism promised in \cref{sec-sub:motiv:assumption-based-parallel-specsyn}. {\verdict} addresses this issue via \emph{assumption-based verification}: The agent constructs provisional \emph{assumed contracts} for any callee with an unpublished contract directly from that callee's body, and proceeds to generate and verify its own specification under these assumptions.

The verification procedure then follows a disciplined cycle: The agent reads the host function's body together with currently published contracts of its callees, hypothesizes contracts for the remaining ones, generates ACSL annotations (including function contracts and intraprocedural specifications), and invokes {\Wp} for targeted verification. It then records the status of each \emph{verification obligation} (VO). Three categories of VOs are tracked: (i) proof goals within the host function (e.g., RTE assertions), (ii) host's specification clauses (postconditions, frame conditions, and loop contracts), and (iii) callee preconditions that must be discharged at each call site.

\subsection{Event-Driven Bidirectional Reconciliation}\label{sec-sub:event-driven-incremental}

After its initial cycle, an agent enters an idle state and will be revived only when the orchestrator delivers a pertinent event. Two types of events drive the bidirectional reconciliation:

\emph{Upward cascading}: An \emph{Event Supply} signals that a callee has published a new contract version. The host agent re-verifies the affected VOs against the updated contract;\footnote{The re-verification can sometimes be avoided by comparing the logical strength of the updated contract against its old version.} if they still hold, the callee's change is compatible with the agent's existing proof, and the agent returns to idle.

\emph{Downward request}: An \emph{Event Demand} carries a structured request from a caller: The caller requires a specific property from the callee to discharge one of its own VOs, and has verified that the property suffices under an assumed contract. The corresponding callee agent assesses whether the requested property is provable given its current implementation; if so, it incorporates the property into its contract and publishes it.

When either event reveals a conflict---the new callee contract invalidates the agent's proof, or the demanded property cannot be established---the agent proceeds with reconciliation. First, it attempts an intraprocedural fix; for instance, a refined loop invariant may suffice to re-establish the altered callee contract or accommodate the demanded property.
When the agent deems such a fix impossible, it has two further options:
(i) \emph{strengthening its precondition} and publishing a revised contract, which in turn triggers Event Supply for its callers and propagates the change upward; and (ii) \emph{issuing demands to its own callees}---requesting a stronger postcondition or a weaker precondition---which triggers Event Demand for those callees and propagates the conflict downward.

Note that these two options are \emph{non-exclusive}: an agent may pursue both, and in either case, the reconciliation generates new events that activate further agents. Thus, although a Supply travels only upward and a Demand only downward, each can trigger the other: a Demand reaching a callee may lead it to strengthen its precondition and raise a Supply to its callers, and vice versa. This interleaving of %
events across the call graph achieves the bidirectional reconciliation.

\subsection{Agent Prompt, Tools, and Loop}\label{sec-sub:agent-loop}

Each agent is instructed by a prompt that specifies its role as the exclusive owner of the host function, its task of publishing the host's contract with all its VOs discharged, the verification cycle of \cref{sec-sub:assumption-based-verification}, and when and how to use each tool.
The agent acts through four domain-specific tools, which realize the steps of \cref{sec-sub:assumption-based-verification,sec-sub:event-driven-incremental}: the \emph{Context} tool reads the code of the host function and its callees, along with their published contracts; the \emph{Verification} tool invokes {\Wp} on candidate annotations and records the status of each VO; the \emph{Supply} tool publishes the host's contract, triggering Event Supply for its callers; and the \emph{Demand} tool requests a stronger postcondition or a weaker precondition from a callee, triggering Event Demand. Generic built-in tools (e.g., Read, Grep, and Bash) are also available.
\agentdetails{}
The agent runs these tools in an adaptive loop: within a cycle, it decides which tools to call, reads their results, and decides when the cycle ends, and each cycle resumes its previous session. When an agent is dispatched, and with which pending events, is decided by the orchestrator, which we present next.

\begin{figure*}[t]
    \centering
    \includegraphics[width=\linewidth]{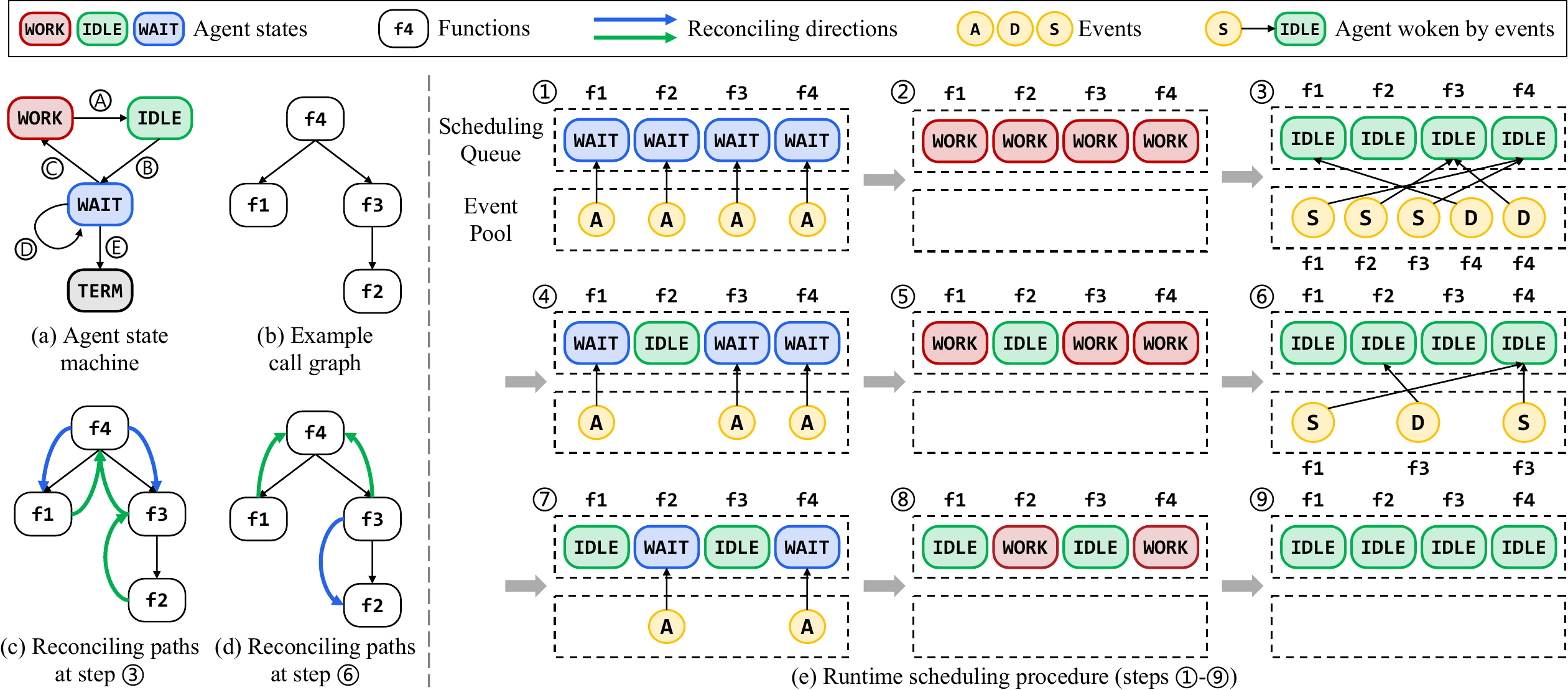}
    \caption{Illustration of {\verdict}'s multi-agent scheduling mechanism. (a)~Agent state machine. (b)~Example call graph. (c)--(d)~Bidirectional reconciliation invoked by events at steps~\Circled{3} and~\Circled{6}, respectively. (e)~Scheduling trace: the processing queue and event pool evolve across nine steps \Circled{1}--\Circled{9} until termination (see \cref{sec-sub:termination}).
    }
\label{fig:runtime-orchestrator}
\end{figure*}

\section{Orchestrator: Multi-Agent Scheduling}\label{sec:orchestrator}

\subsection{Agent State Machine}\label{sec-sub:agent-state-machine}

To coordinate the concurrent execution of all agents, the orchestrator models each agent as a four-state machine, as shown in \cref{fig:runtime-orchestrator}~(a). The states of an agent $\mathcal{A}$ capture its progress throughout the verification lifecycle:
\begin{itemize}[leftmargin=*, itemindent=0pt, label=--]
\item \texttt{WORK}: $\mathcal{A}$ actively verifies its specification against the current version of its callee contracts (cf. \cref{sec-sub:assumption-based-verification,sec-sub:event-driven-incremental}); 
\item \texttt{IDLE}: $\mathcal{A}$ has completed its current cycle and awaits events; 
\item \texttt{WAIT}: $\mathcal{A}$ has received an event and is queued for execution; 
\item \texttt{TERM}: $\mathcal{A}$ has permanently exited the scheduling queue.
\end{itemize}

The orchestrator governs five transitions between these states. When agent $\mathcal{A}$ finishes its verification cycle, it moves from \texttt{WORK} to \texttt{IDLE} (transition~\Circled{A}). The orchestrator then inspects whether any event (Supply or Demand) is pending for $\mathcal{A}$; if so, it attaches the event to $\mathcal{A}$ and advances $\mathcal{A}$ to \texttt{WAIT} (transition~\Circled{B}).
When a slot in the \emph{concurrency pool}---which caps the number of agents that may run simultaneously; set by hyper-parameter \texttt{max\_parallel}---becomes available, the orchestrator \emph{dispatches} a \texttt{WAIT} agent into \texttt{WORK} (transition~\Circled{C}). If the pool is full, $\mathcal{A}$ remains in state \texttt{WAIT} (transition~\Circled{D}).
Finally, if $\mathcal{A}$ has exhausted its \emph{cycle budget}---the maximum number of \texttt{WORK} rounds it is permitted to perform; set by hyper-parameter \texttt{max\_cycles}---it transitions from \texttt{WAIT} to \texttt{TERM} (transition~\Circled{E}), permanently exiting the scheduling queue.

\subsection{Scheduling Procedure}\label{sec-sub:scheduling-procedure}

\Cref{fig:runtime-orchestrator}~(e) depicts the scheduling procedure on the call graph of \cref{fig:runtime-orchestrator}~(b) across nine steps, with $\texttt{max\_parallel}=4$. Agents are enqueued in reverse topological order of the call graph such that, under limited concurrency, agents deeper in the call chain are dispatched first---intuitively, resolving callees before callers reduces the number of revivals. We use this concrete example to illustrate how the state machine and the events interact to drive convergence.

\paragraph*{\bf Initial dispatch (steps \Circled{1}--\Circled{2}).} 
All four agents are launched concurrently and enter \texttt{WAIT}, each carrying an \emph{Event Available} (denoted as \texttt{A} in \cref{fig:runtime-orchestrator})---a special signal from the orchestrator that a concurrency-pool slot has been assigned. When the pool admits all four, they transition to \texttt{WORK} and execute their first assumption-based verification cycle.

\paragraph*{\bf First reconciliation (steps \Circled{3}--\Circled{5}).} 
After all agents return to \texttt{IDLE}, the orchestrator attaches events. In the event pool of \Circled{3}, each label below an event indicates the source agent who triggers the event, and each arrow points to the target agent that will be revived, as per \cref{fig:runtime-orchestrator}~(c). 
Intuitively, \texttt{f1}/\texttt{f2}/\texttt{f3} publish new contracts (upward Supply), while \texttt{f4} demands property adjustments from \texttt{f1} and \texttt{f3} (downward Demand). In \Circled{4}, \texttt{f2}---whose verification has no conflicts and receives no events---remains \texttt{IDLE}. The others enter \texttt{WORK} in \Circled{5}.

\paragraph*{\bf Second reconciliation (steps \Circled{6}--\Circled{9}).} 
After the three agents complete, new events emerge as shown in \cref{fig:runtime-orchestrator}~(d): two Event Supply from \texttt{f1} and \texttt{f3} target \texttt{f4}; one Event Demand from \texttt{f3} targets \texttt{f2}. In \Circled{7}, \texttt{f2} and \texttt{f4} enter \texttt{WAIT} while \texttt{f1} and \texttt{f3} remain \texttt{IDLE}. After \texttt{f2} and \texttt{f4} complete in \Circled{8}, no further events are generated. Upon step \Circled{9}, \emph{all agents are} \texttt{IDLE} \emph{with an empty event pool}---the \emph{orchestrator terminates.}

\begin{table*}[t]
\centering
\caption{Specification quality and verification time for RQ1. \textbf{\#Fn}/\textbf{Dep}/\textbf{LoC}: functions, call-chain depth, lines of code (of the \framac processed program, which may be larger than the original source). For each tool, \textbf{rte}/\textbf{spec} are the percentages of runtime-error and LLM-synthesized goals discharged by {\Wp}, and \textbf{time} is end-to-end minutes, each as mean$\pm$std over three repetitions. \emph{Overall} is the rte/spec rate over all benchmarks' goals and the geometric-mean time.
\textbf{speedup} is \verdict's factor over \preguss and SpecSyn (either \textcolor{ForestGreen}{faster} or \textcolor{gray}{slower}).}
\label{tab:rq1}
\includegraphics[width=\textwidth]{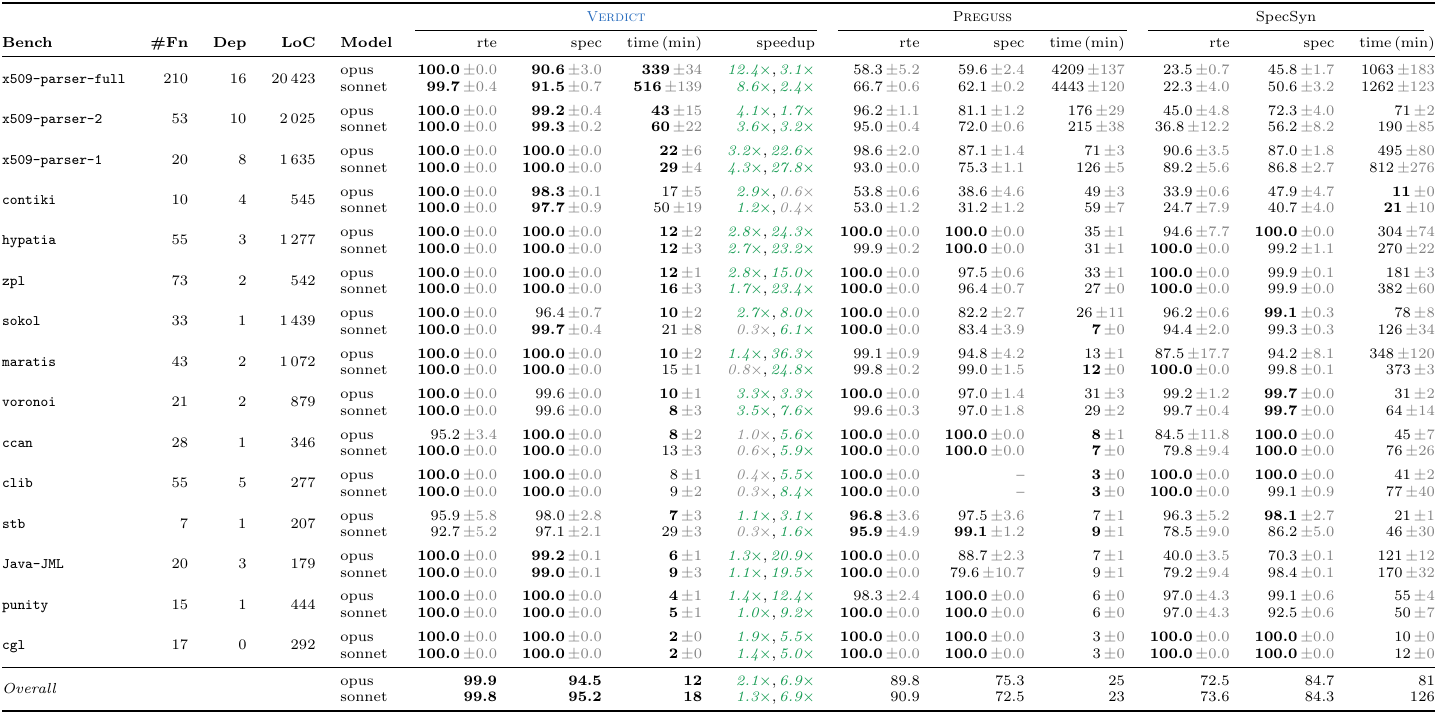}
\end{table*}

\subsection{Termination of the Orchestrator}\label{sec-sub:termination}

The orchestrator \emph{terminates} under either of two conditions: (1)~the scheduling queue contains only \texttt{IDLE} or \texttt{TERM} agents and the event pool is empty;
or (2)~all agents have transitioned to \texttt{TERM}, at which point remaining events are discarded since no agent remains to consume them. Termination is guaranteed by construction. Each agent is subject to a \emph{cycle budget} %
and each \texttt{WORK} round runs within a finite time limit. Consequently, every agent eventually either settles into \texttt{IDLE} with no pending events or exhausts its budget and stays permanently in \texttt{TERM}.

\subsection{Soundness of the Orchestrator}\label{sec-sub:soundness}

An agent \emph{converges} if the following conditions hold upon termination of the orchestrator: (i)~it rests in \texttt{IDLE}, (ii)~every obligation it tracks is discharged, and (iii)~no pending event targets it. To demonstrate soundness,
let $G=(V,E)$ be the call graph of the program, with each $v\in V$ a function and each $(v_i,v_j)\in E$ a call from $v_i$ to $v_j$. For any subgraph $H\subseteq G$, denote by $\mathcal{A}_H$ the set of agents responsible for the functions in $H$. Let $V_0\subseteq V(H)$ be the entry functions---those not called by any other function in $H$.
Then, the orchestrator is \emph{sound} in the following sense: For all $H\subseteq G$, whenever all agents in $\mathcal{A}_H$ converge, the proof goals of $H$ are valid if the preconditions of every function in $V_0$ are satisfied.

\section{Evaluation}\label{sec:evaluation}

We implement {\verdict} and evaluate it by addressing the following research questions (RQs): 

\begin{itemize}
\item \textbf{RQ1}: How does \verdict compare against state-of-the-art LLM-based specification-synthesis approaches?
\item \textbf{RQ2}: What are the effects of \verdict's scheduling hyper-parameters (\cref{sec-sub:agent-state-machine}) on its performance?
\item \textbf{RQ3}: To what extent do the two mechanisms (\cref{sec-sub:assumption-based-verification,sec-sub:event-driven-incremental}) of {\verdict} contribute to its effectiveness?
\end{itemize}

\subsection{Experimental Setup}\label{sec-sub:eval:setup}

\paragraph*{Implementation.}
\verdict is implemented atop the Claude Agent SDK~\cite{claude-agent-sdk}\NEWCITE{} with \texttt{asyncio}-based concurrency. Each agent invokes {\Wp}~\cite{frama-c/wp} with the SMT solvers Z3~\cite{z3}\NEWCITE{}, cvc5~\cite{cvc5}\NEWCITE{}, and Alt-Ergo~\cite{alt-ergo}\NEWCITE{} for deductive verification.

\paragraph*{Models and configurations.}
Since \verdict is built on the Claude Agent SDK, we employ two Claude models: Sonnet~4.6~\cite{claude-sonnet-4-6}\NEWCITE{} and Opus~4.5~\cite{claude-opus-4-5}\NEWCITE{}.
All LLM sampling parameters are left at their API defaults. Prior to the evaluation, we fixed the hyper-parameters $\texttt{max\_parallel}=8$ and $\texttt{max\_cycles}=24$, except for benchmark \texttt{x509-parser-full} where we use $\texttt{max\_parallel}=16$ given its scale.
We repeat all RQ1 experiments three times and report the mean to account for LLM non-determinism. For RQ2 and RQ3, we run a single trial with both models for cross-validation.
All experiments are conducted on an Ubuntu~24.04 server with 2 AMD EPYC~7542 processors (128~threads in total) and 120~GiB of RAM.

\paragraph*{Baselines.}
We compare \verdict against two state-of-the-art LLM-based specification-synthesis tools---\preguss~\cite{preguss} and SpecSyn~\cite{SpecSyn}, as detailed in \cref{sec:related-work}---which can handle programs with non-trivial call graphs by adopting the divide-and-conquer, bottom-up paradigm.
Both baselines use the same LLMs as \verdict to ensure a fair comparison.

\paragraph*{Benchmarks.}
We assemble our benchmark set from the evaluation corpora of \preguss and SpecSyn.
Specifically, we omit benchmarks of trivial size, e.g., Frama-C-Problems and SV-COMP, as well as \texttt{atomthreads}, whose complex data structures (e.g., linked lists) require inductive predicates and thus lie beyond the scope of {\verdict} and the baselines.
The benchmark set is supplemented by two additional modules extracted from the ANSSI X.509 certificate parser~\cite{x509-parser}.
As collected in \Cref{tab:rq1}, the benchmarks vary widely in call-chain depth (0 to 16) and program size (7 to 210 functions). 

\paragraph*{Metrics.}
We run {\Wp}---under the same configuration with consistent memory model, solver timeout (30\,s), and prover set---on each tool's output (programs with annotated specifications) and report the \emph{fraction of discharged proof goals} in two categories: \emph{rte} (runtime-error goals generated by \rte~\cite{frama-c/rte}) and \emph{spec} (LLM-synthesized loop contracts, postconditions, frame conditions, and callee-precondition obligations at call sites). We additionally report end-to-end wall-clock time.

\subsection{RQ1: Comparison against Baselines}\label{sec-sub:eval:rq1}

\Cref{tab:rq1} reports the experimental results of {\verdict} against {\preguss} and SpecSyn. In general, we observe two distinct advantages of {\verdict} over these baselines:

\paragraph*{\bf Specification quality.}
Across both models, \verdict \emph{discharges the largest fraction of proof goals in both categories}. Pooled over all goals (Opus), it reaches 99.9\% on rte and 94.5\% on spec, against 89.8\%/75.3\% for \preguss and 72.5\%/84.7\% for SpecSyn, with consistent Sonnet results. In particular, \verdict discharges 100\% runtime-error goals on 12 of the 15 benchmarks under both models, compared with 6 for \preguss and 3 for SpecSyn. The two baselines also reflect their respective design emphases: \preguss, being RTE-guided, scores markedly higher on rte than on spec; whereas SpecSyn, which optimizes specification strength, exhibits the opposite skew. Notably, \verdict \emph{surpasses each baseline on its own stronger dimension}, i.e., \preguss's rte and SpecSyn's spec. 

\verdict's lead is most pronounced for programs with \emph{deep call chains}, where interprocedural specifications are hard to synthesize. For instance, on the full X.509 parser with depth 16, conflicts must be reconciled across many call levels, and the baselines drop to 58.3\%/59.6\% (rte/spec) for \preguss and 23.5\%/45.8\% for SpecSyn, while \verdict achieves 100\%/90.6\%.
This advantage stems from the bidirectional reconciliation mechanism, which we isolate in RQ3.
On shallow programs dominated by local reasoning, e.g., \texttt{cgl} and \texttt{zpl}, the three tools are far closer, all approaching 100\%.

\paragraph*{\bf Verification time.}
With Opus, \verdict is the most efficient tool on 12/15 benchmarks, with a geometric-mean time of 12 minutes, compared to 25 for \preguss and 81 for SpecSyn. Moreover, \verdict's \emph{advantage increases with program size and call-chain depth:} Its modest $1.4{\times}$ speedup over \preguss on small, shallow programs (\texttt{punity}; \text{\#Fn}=15, \text{Dep}=1) grows to $2.8{\times}$ on larger yet still shallow ones (\texttt{hypatia}; \text{\#Fn}=55, \text{Dep}=3), reaches $4.1{\times}$ with deep calls (\texttt{x509-parser-2}; \text{Dep}=10), and culminates at $12.4{\times}$ on \texttt{x509-parser-full} (\text{\#Fn}=210, \text{Dep}=16; 339 vs.\ 4{,}209 min). In exchange, \verdict makes more LLM calls; its cost relative to \preguss is higher on smaller programs and drops to $2.7{\times}$ (Opus) and $3.1{\times}$ (Sonnet) on \texttt{x509-parser-full}. 

\begin{tcolorbox}[colback=gray!5, colframe=gray!50, boxrule=0.5pt, sharp corners, left=4pt, right=4pt, top=2pt, bottom=2pt]
\textbf{Answer to RQ1:}
\verdict substantially outperforms state-of-the-art baselines in both specification quality and efficiency.
In particular, \verdict is the first automated approach capable of proving the near-complete absence of runtime errors for the entire 20{,}000-line X.509 parser in a few hours, whereas its original manual verification required months of expert effort~\cite{x509-parser}.
\end{tcolorbox}

\subsection{RQ2: Effects of the Scheduling Parameters}\label{sec-sub:eval:rq2}

\verdict exposes two hyper-parameters (\cref{sec-sub:agent-state-machine}): \texttt{max\_parallel}~($mp$) and \texttt{max\_cycles}~($mc$). We evaluate their effect on six representative benchmarks (see \cref{fig:rq2a}) spanning shallow to deep call chains.

\subsubsection{\texttt{max\_parallel}.}\label{sec-subsub:eval:rq2a}

\begin{figure}[t]
\centering
\includegraphics[width=\linewidth]{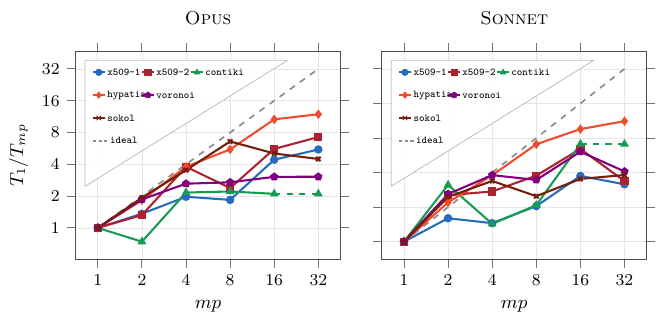}
\caption{Parallel speedup $T_1/T_{mp}$ vs.\ $mp$ ($mc=24$), per model. The dashed line marks the ideal $T_1/T_{mp}{=}mp$; \texttt{contiki} has only 10 functions, so its $mp{=}32$ point repeats $mp{=}16$. 
}
\label{fig:rq2a}
\end{figure}

We fix $mc=24$ and vary $mp \in \{1, 2, 4, 8, 16, 32\}$. When $mp=1$, \verdict reduces to the sequential bottom-up order, verifying one function at a time.
We measure the gain as the \emph{parallel speedup} $T_1/T_{mp}$, i.e., the wall-clock time at $mp=1$ divided by the time with $mp$ agents. As depicted in \cref{fig:rq2a}, the speedup grows steadily along $mp$: with Opus, \texttt{hypatia} rises from 1.8$\times$ at $mp{=}2$ to 11.9$\times$ at $mp{=}32$, and \texttt{x509-parser-2} falls from 150 to 21 min, a 7$\times$ gain; Across the 6 benchmarks, the geometric-mean speedup at $mp{=}32$ is 4.9$\times$. Sonnet yields similar observations.

\begin{figure}[t]
\centering
\includegraphics[width=\linewidth]{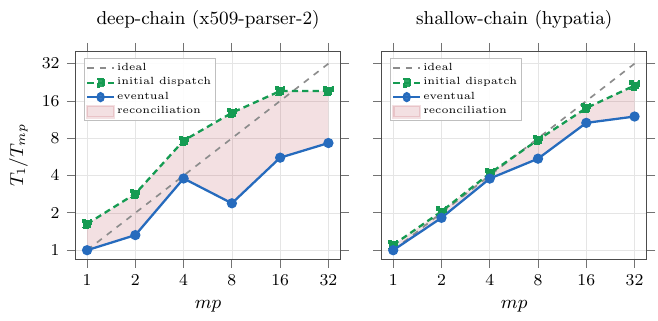}
\caption{Speedup decomposition (\textsc{Opus}). For each program, the \emph{eventual} speedup $T_1/T_{\mathrm{total}}$ (the parallel speedup of \cref{fig:rq2a}) and the \emph{initial-dispatch} speedup $T_1/T_{\mathrm{init}}$ are plotted against $mp$, and the gap between them is the speedup lost to \emph{reconciliation}.
}
\label{fig:rq2c}
\end{figure}

Note that these parallel speedups are \emph{sub-linear}: all curves exhibit only gentle growth and remain below the ideal scaling of $T_1/T_{mp}{=}mp$. \texttt{hypatia} (\text{Dep}=3) stays closest to this bound, with a slope of roughly three-quarters of the ideal, whereas \texttt{x509-parser-2} (\text{Dep}=10) increases at barely half that rate.
This gap stems from the cost of interprocedural reconciliation.
With few agent slots, the schedule remains close to the dependency order: a callee typically publishes a contract before its callers start, so little reconciliation is required. As $mp$ increases, more agents start working during the \emph{initial dispatch} (\cref{sec-sub:scheduling-procedure}) before their callees have published contracts, leaving those assumed contracts to be resolved afterward through \emph{reconciliation}. Thus, we have
\begin{align*}
\stackunder{T_{\mathrm{total}}}{\overbrace{\substack{\textnormal{total time}}}} \eeq \stackunder{T_{\mathrm{init}}}{\overbrace{\substack{\textnormal{time for initial dispatch}}}} + \stackunder{T_{\mathrm{rec}}}{\overbrace{\substack{\textnormal{time for reconciliation}}}}
\end{align*}%
where the reconciliation overhead $T_{\mathrm{rec}}$ grows with increasing parallelism.
\Cref{fig:rq2c} isolates this effect on two benchmarks with distinct call-chain depths. On \texttt{x509-parser-2} (Dep=10), the gap between the initial-dispatch and eventual curves is substantial: the initial-dispatch potential reaches 19$\times$ at $mp{=}32$, but reconciliation drags the eventual speedup back to 7$\times$. On \texttt{hypatia} (Dep=3), where functions exhibit limited interdependence, the gap remains much narrower and the eventual speedup climbs to 12$\times$. This reconciliation is thus not wasted effort but \emph{the price of correctness at speed}: \verdict spends it resolving interprocedural conflicts that ripple through deep call chains (see \cref{sec-sub:motiv:bid-interpro-conflict-reconcile}).

\subsubsection{\texttt{max\_cycles}.}\label{sec-subsub:eval:rq2b}

The cycle budget $mc$ caps how many refinement rounds each agent may perform.
With $mp=8$ and $mc \in \{1, 2, 4, 8, 16, 24\}$, we track two rates: the external \emph{discharge rate}, i.e., the fraction of rte and spec goals verified independently by {\Wp}; and the internal \emph{convergence rate}, 
i.e., the fraction of converged agents (\cref{sec-sub:soundness}).

\begin{figure}[t]
\centering
\includegraphics[width=\linewidth]{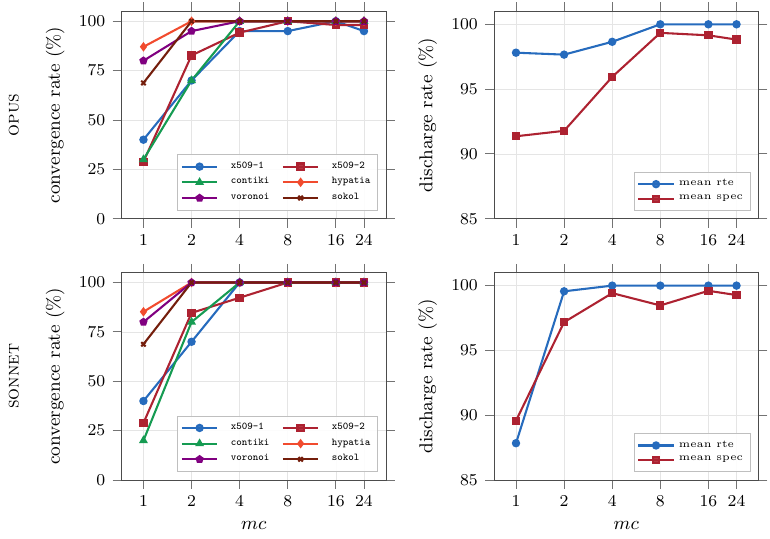}
\caption{Convergence rate (left) and mean rte/spec discharge (right) vs.\ $mc$ ($mp=8$); rows are models. 
}
\label{fig:rq2b}
\end{figure}

\Cref{fig:rq2b} depicts both rates against $mc$. The convergence rate rises sharply from a low base: the deep-chain X.509 submodules sit under half at $mc{=}1$ and top out by $mc{\approx}8$, whereas the shallow-chain benchmarks reach saturation by $mc{\approx}2$ to 4. This indicates that the number of refinement rounds required grows with call-chain depth. This is the same depth-driven reconciliation effect observed in \cref{fig:rq2c}, manifested there in wall-clock time and here in refinement iterations. The externally verified discharge rate improves on the same schedule, with mean spec discharge increasing from ${\approx}90\%$ to ${\approx}99\%$ by $mc{\approx}8$ under both models. Internal convergence and external quality thus plateau in tandem: \verdict's notion of completion closely tracks the quality certified by {\Wp}.

\begin{tcolorbox}[colback=gray!5, colframe=gray!50, boxrule=0.5pt, sharp corners, left=4pt, right=4pt, top=2pt, bottom=2pt]
\textbf{Answer to RQ2:}
\verdict's two hyper-parameters govern complementary axes: more agents improve efficiency, while more cycles improve quality.
\end{tcolorbox}

\subsection{RQ3: Contribution of Individual Mechanisms}\label{sec-sub:eval:rq3}

We perform an ablation study to quantify each mechanism's individual contribution to \verdict's specification quality and efficiency.
Specifically, we evaluate four configurations across the six RQ2 benchmarks: the full system, \emph{no-assumption} ($-$asm, an agent can rely only on a callee's published contract, not on an assumed one), \emph{no-demand} ($-$dem, an agent can no longer raise an \emph{Event Demand})\footnote{Event Supply is not ablated to retain soundness of verification: It triggers an agent to re-verify its proof whenever a callee publishes a new contract.}, and \emph{no-both} ($-$both).

\begin{figure}[t]
\centering
\includegraphics[width=\linewidth]{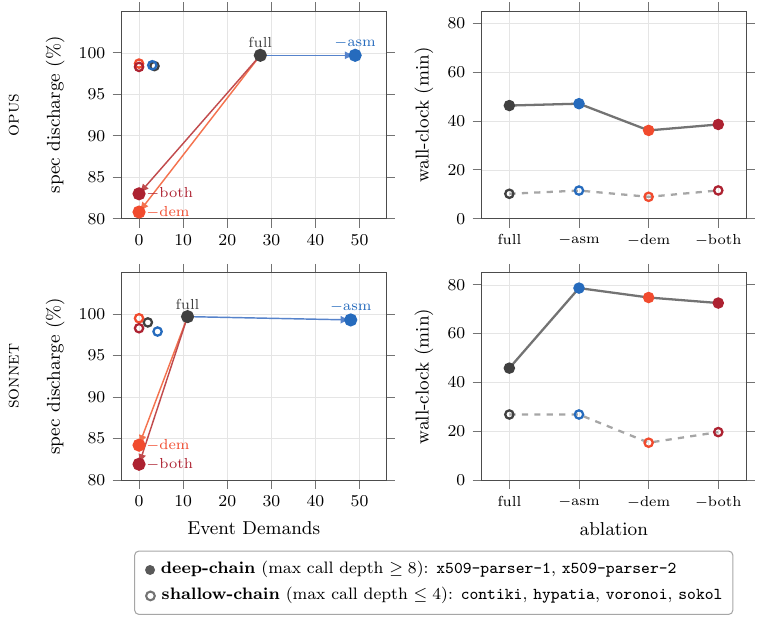}
\caption{Ablation on the six RQ2 benchmarks, averaged per call-chain-depth group. Left: Specification discharge against the number of Event Demands, with arrows from the full system to each ablation ($-$asm, $-$dem, $-$both); Right: Wall-clock time.}
\label{fig:rq3}
\end{figure}

The results split cleanly by call-chain depth; see \cref{fig:rq3}.
On the four shallow programs (\texttt{contiki}, \texttt{hypatia}, \texttt{voronoi}, \texttt{sokol}; Dep $\leq 4$), all configurations cluster in the top-left corner with near-perfect quality, so ablation barely registers. The cost is concentrated in the two X.509 submodules (depth 8 and 10), where we analyze each mechanism in turn.

\paragraph*{Removing Event Demand reduces quality.} 
Under no-demand, the deep-chain point moves down the quality axis: Specification discharge rate drops from 99.7\% to 81\% (Opus) and 84\% (Sonnet). The no-both configuration yields similar outcomes. In fact, without Event Demand, an agent can no longer ask a callee to repair a cross-level conflict (Conflict~A or~B in \cref{sec-sub:motiv:bid-interpro-conflict-reconcile}), exposing precisely the failure mode that the bottom-up paradigm cannot avoid.

\paragraph*{Removing assumed contracts reduces efficiency.} 
Under no-assumption, the deep-chain point instead shifts rightward, maintaining quality above 99\% while the number of Demand Events increases sharply, from 28 (11) to 49 (48) with Opus (Sonnet). Without optimistic assumptions about callees that have not yet published, each dependency must be resolved sequentially via Event Demands, eliminating the parallelism that assumed contracts are designed to enable (\cref{sec-sub:motiv:assumption-based-parallel-specsyn}).

\begin{tcolorbox}[colback=gray!5, colframe=gray!50, boxrule=0.5pt, sharp corners, left=4pt, right=4pt, top=2pt, bottom=2pt]
\textbf{Answer to RQ3:}
The two mechanisms address complementary objectives: Event Demand safeguards quality, while assumed contracts deliver efficiency. On programs with deep, intricate call structures that the bottom-up paradigm cannot adequately handle, neither mechanism can be removed without loss. At the same time, their neutrality on shallow programs reinforces the same point from the other side: \verdict engages reconciliation only where conflicts arise, and incurs no overhead where they do not.
\end{tcolorbox}

\subsection{Threats to Validity}\label{sec-sub:threats}

We measure specification quality as the fraction of proof goals discharged by a unified, external {\Wp} configuration.
A more rigorous alternative, as adopted in \preguss~\cite{preguss}, would be to manually refine the generated specifications until 100\% discharge is achieved, and then measure the human effort required. We opted against this for two reasons:
(i)~it is prohibitively expensive on the benchmarks where the comparison is most informative---on \texttt{x509-parser-full}, both baselines leave a substantial number of goals undischarged, making manual validation significantly more labor-intensive than on smaller programs; and
(ii)~even on the subset of benchmarks that already achieve 100\% discharge---where no human intervention is required---\verdict still leads (12 of 15 on rte, compared to 6 for \preguss and 3 for SpecSyn). The discharge rate is therefore a reasonable metric, since a tool that discharges more goals leaves arguably less work for subsequent manual refinement.

\section{Limitations and Future Directions}\label{sec:limitations}

{\verdict}'s design is not yet fully exploited in its current instantiation; we outline three directions it opens up at the levels of scheduling, architecture, and specification.

\paragraph*{Overlapping reconciliation rounds.}
RQ2 has identified reconciliation---the rounds in which each agent resolves its optimistic assumptions after its callees publish---as the primary source of the gap between eventual and ideal speedup (\cref{fig:rq2c}). {\verdict} currently executes these reconciliation rounds in a synchronous manner (\cref{sec-sub:scheduling-procedure}), completing one round before starting the next. Allowing a revived agent to resume immediately upon receiving its triggering event, rather than waiting for the next round boundary, would enable overlap across consecutive rounds, thereby reducing this tail effect and bringing the eventual speedup closer to the ideal.

\paragraph*{Rethinking the granularity of agents.} 
\verdict assigns one agent to every function (\cref{sec-sub:motiv:assumption-based-parallel-specsyn}).
This choice maximizes parallelism and keeps each agent's context small, thereby mitigating the long-context reasoning bottleneck. Because agents proceed under assumed contracts and later reconcile them through events, even cyclic dependencies that admit no bottom-up order are handled at this level of granularity: mutually recursive functions within a \emph{strongly connected component} (SCC) of the call graph resolve their dependencies via Supply and Demand events until their contracts converge. The cost of this design is that every interprocedural dependency becomes a cross-agent negotiation mediated by assumed contracts and proof demands. The functions that are most tightly coupled are therefore also those that trigger the most renegotiation, contributing disproportionately to tail latency. This raises a natural question of whether function-level granularity is the right choice. An alternative is to assign a single agent to a \emph{region of tightly coupled functions}, such as an entire SCC. Such an agent could discharge internal dependencies locally, invoking assumed contracts and proof demands only across region boundaries. In general, an appropriate granularity must balance three forces: parallelism, which favors finer units; coordination overhead, which favors coarser units; and per-agent context size, which grows with region size. How to partition the call graph to strike this balance remains open.

\paragraph*{Synthesizing richer specifications.} 
\verdict, like the baselines, targets the absence of RTEs alongside the contracts, loop invariants, and frame conditions that establish it. Programs with recursive, unbounded data structures, e.g., linked lists and trees, additionally require inductive predicates and lemmas, which remain challenging for current LLM-based synthesis. Equipping agents to construct such artifacts is a natural way to broaden the class of properties that \verdict can certify.

\section{Related Work}\label{sec:related-work}

LLM-based specification synthesis~\cite{cav24-autospec,icse25-specgen,preguss,SpecSyn,move,oopsla25-autoverus,oopsla25-LLM4assertion,tacas26-veristruct} has undergone a recent surge of interest due to its potential for automating deductive verification~\cite{specification-synthesis}. These approaches adopt a common \emph{guess-and-check} paradigm: an LLM proposes candidate specifications (e.g., function contracts, loop invariants, assertions), a deductive verifier checks them, and verification failures are fed back for iterative refinement.

Existing approaches can be broadly divided by scale. The majority target \emph{intraprocedural} or \emph{single-function} settings: AutoSpec~\cite{cav24-autospec} and SLD-Spec~\cite{sld-spec} decompose programs into extended call graphs with mask-based placeholder substitution; SpecGen~\cite{icse25-specgen} applies mutations to per-method JML~\cite{jml} specifications; Laurel~\cite{oopsla25-LLM4assertion} localizes missing Dafny~\cite{dafny} assertions from verifier error messages; LORIS~\cite{toplas26-loris} improves loop invariant synthesis by formally justifying the LLM's natural-language reasoning steps; AutoVerus~\cite{oopsla25-autoverus} and Msg~\cite{move} employ multi-agent architectures for Rust/Verus and Move, respectively, but still operate one function at a time. Although these systems achieve impressive per-function success rates, they are limited to functions with shallow or no call dependencies.

Two systems stand out for tackling \emph{interprocedural} specification synthesis \emph{at scale}:  {\preguss}~\cite{preguss} decomposes programs into V-Units---a static, two-layer program context. 
SpecSyn~\cite{SpecSyn} partitions programs into SCCs together with their necessary dependencies, yielding similarly small, locally bounded units.
Both approaches adopt a divide-and-conquer, bottom-up strategy: restricting each unit avoids long-context reasoning, while verifying callees before callers attains soundness.
Such a strategy, however, imposes fundamental limitations as detailed in \cref{sec:motivation}. A verification unit can refine only the specifications of functions within its static boundary; influencing functions beyond that boundary requires restarting the entire bottom-up process. 
Consequently, neither system can propagate specification demands across multiple call levels, preventing the interleaved conflict reconciliation as presented in \cref{sec-sub:motiv:bid-interpro-conflict-reconcile}.
\verdict addresses these limitations through the mechanisms of assumption-based verification and bidirectional interprocedural conflict reconciliation (\cref{sec:agent}).

FM-Agent~\cite{fm-agent} also departs from the bottom-up paradigm, but in the opposite direction: It generates specifications \emph{top-down}, inferring each callee's contract from its callers and using LLM-based Hoare-style reasoning to \emph{detect bugs} against natural-language specifications. FM-Agent is not our baseline since it targets bug finding rather than deductive verification.

\section{Conclusion}\label{sec:conclusion}

We have presented \verdict, a multi-agent framework that parallelizes modular deductive verification by replacing the sequential bottom-up paradigm with assumption-based parallel dispatch and event-driven bidirectional reconciliation. 
We showcase that \verdict substantially outperforms state-of-the-art baselines in both specification quality and efficiency, and enables the almost fully automated RTE-freeness verification of real-world programs with over 20,000 LoC, which has previously been limited to the size of 1,000 LoC.

\section{Data Availability}\label{sec:data-availability}

The benchmarks, experimental results, Docker images, and evaluation scripts are publicly available at \url{https://doi.org/10.5281/zenodo.24453472}.

\ifshowacknowledgment
\section*{Acknowledgment}
This work has been partially supported by the NSFC (No.\ 62572427), by the ZJNSF Major Program (No.\ LD24F020013), by the Huawei Cooperation Project (No.\ TC20260720069), by the CASC Open Fund (No.\ LHCESET202502), by the CCF-Huawei Populus Grove Fund (No.\ CCF-HuaweiSY202503), and by the Zhejiang Key Laboratory Project (No.\ 2024E10001).
\fi

\bibliographystyle{IEEEtran}
\bibliography{IEEEabrv,References}

\onecolumn %
\appendices
\crefalias{section}{appendix}%
\newcommand{\aptool}[1]{\texttt{#1}}
\newcommand{\apslot}[1]{$\langle$\textit{#1}$\rangle$}

\section{Agent Prompt and Tools}\label{app:agent}

Each agent is instructed by the prompt below (\cref{sec-sub:agent-loop}), instantiated for its host function, and each later cycle opens with a cycle message that carries the events that revived the agent and the status of its VOs. \Cref{tab:tools} lists the tools.

\begin{promptbox}{Agent prompt}
\textbf{Role.} You are the exclusive owner of the host function \apslot{f}.\\
\textbf{Host function:} \apslot{source code of f}.\quad \textbf{Callees:} \apslot{names of callees}.\quad \textbf{Callers:} \apslot{names of callers}.

\smallskip
\textbf{Task.} Synthesize a verified ACSL contract for \apslot{f}, discharge all its verification obligations (VOs: RTE, specification, and call-site), and publish the contract. Coordinate with other agents only through Supply and Demand events.

\smallskip
\textbf{Tools.} \apslot{for each tool in \cref{tab:tools}: when and how to use it}

\smallskip
\textbf{Verification cycle.}\par\smallskip
\noindent\begin{tabularx}{\linewidth}{@{}r@{\hspace{4pt}}X@{}}
1. & Read the host's code and the published contracts of its callees. \\
2. & Generate ACSL annotations and verify them with {\Wp} under the published callee contracts, assuming the missing or insufficient ones. \\
3. & Discharge the remaining VOs, then publish the contract, which sends a Supply to the callers. \\
4. & When revived by a Supply, re-verify the affected VOs; when revived by a Demand, accept it if the demanded property is provable. If a conflict remains, first try an intraprocedural fix; otherwise strengthen the host's precondition and republish, or raise a Demand on a callee for a stronger postcondition or a weaker precondition. Reject a Demand only if all of these fail. \\
\end{tabularx}
\end{promptbox}

\begin{promptbox}{Cycle message}
\textbf{Revived by:} \apslot{Supply from callee g: new contract} \;$\mid$\; \apslot{Demand from caller h: property P}\\
\textbf{VO status:} \apslot{the status of each VO}
\end{promptbox}

\begin{table}[!h]
\centering
\caption{Tools of a {\verdict} agent, grouped as in \cref{sec-sub:agent-loop}.}
\label{tab:tools}
{\small
\begin{tabularx}{\linewidth}{@{}l l X@{}}
\toprule
\textbf{Category} & \textbf{Tool} & \textbf{Purpose} \\
\midrule
Context & \aptool{get\_function\_context(k, include\_callers)} & Code of the host and its callees up to depth \aptool{k} (default 1, i.e., the V-Unit), optionally with its direct callers. \\
 & \aptool{read\_function\_spec(g)} & The latest published contract of a function \aptool{g}, typically a callee. \\
\midrule
Verification & \aptool{verify\_function\_contract(spec, assumed)} & Run {\Wp} on a candidate \aptool{spec}, using \aptool{assumed} callee contracts where published ones are missing or insufficient. \\
 & \aptool{verify\_rte\_assertions(VOs)} & Run {\Wp} on the given RTE VOs. \\
 & \aptool{verify\_callsite\_preconditions(g)} & Run {\Wp} on the call-site VOs of callee \aptool{g}. \\
\midrule
Supply & \aptool{update\_spec} & Publish the verified contract, which sends a Supply to the callers. \\
\midrule
Demand & \aptool{submit\_proof\_demand(g, P)} & Ask callee \aptool{g} for property \aptool{P}: a stronger postcondition or a weaker precondition. \\
 & \aptool{resolve\_proof\_demand(d)} & Accept or reject a Demand \aptool{d} from a caller. \\
\midrule
Generic & Read, Grep, Bash & Built-in tools of the agent SDK. \\
\bottomrule
\end{tabularx}}
\end{table}

\end{document}